\documentclass[twocolumn, twocolappendix]{aastex631}
\usepackage{amsmath}
\usepackage{float}
\usepackage{graphicx}
\usepackage{import}

\usepackage[utf8]{inputenc}
\usepackage{color}
\usepackage{enumitem}

\newcommand{\msun}{M_{\odot}}
\newcommand{\mtot}{M_{\rm tot}}
\newcommand{\mtov}{M_{\rm TOV}}
\newcommand{\mvl}{M_{\rm VL}}
\newcommand{\mls}{M_{\rm LS}}
\newcommand{\mpc}{M_{\rm PC}}
\newcommand{\rsgrb}{R_{\rm sGRB}}
\newcommand{\rlgrb}{R_{\rm lGRB}}
\newcommand{\rgwsgrb}{R_{\rm GW,sGRB}}
\newcommand{\rgwlgrb}{R_{\rm GW,lGRB}}

\begin{document}
\title{Inferring Neutron Star Nuclear Properties from \\Gravitational-Wave and Gamma-Ray Burst Observations}

\correspondingauthor{Hsin-Yu Chen}
\email{hsinyu@austin.utexas.edu}

\author[0000-0001-5403-3762]{Hsin-Yu Chen} 
\affiliation{Department of Physics, The University of Texas at Austin, 2515 Speedway, Austin, TX 78712, USA}
\author[0000-0003-3115-2456]{Ore Gottlieb} 
\affiliation{Center for Computational Astrophysics, Flatiron Institute, New York, NY 10010, USA}
\affiliation{Department of Physics and Columbia Astrophysics Laboratory, Columbia University, Pupin Hall, New York, NY 10027, USA}

\begin{abstract}

Recent discoveries of long gamma-ray bursts accompanied by kilonova emission prompted interest in understanding their progenitors. If these long-duration bursts arise from neutron star mergers, similar to short gamma-ray bursts, it raises the question of which physical properties govern burst duration. The mass of the merger stands out as a key factor, strongly influencing the lifetime of the merger remnant, which in turn determines the burst duration: lighter mergers that form long-lived remnants produce short bursts, whereas more massive mergers result in short-lived remnants that collapse into black holes, powering longer bursts. In this paper, we compare merger rates from gravitational-wave observations of LIGO-Virgo-KAGRA with the rates of kilonova-associated long and short gamma-ray bursts, to identify a characteristic total neutron star mass that separates the two burst classes at $1.36^{+0.08}_{-0.09}$ times of the neutron star Tolman-Oppenheimer-Volkoff (TOV) mass (median and 68\% confidence interval). This result suggests that massive neutron stars could survive an extended period after merger. Our findings are robust against substantial observational uncertainties and model assumptions.  Moreover, {we identify a correlation between the characteristic mass and the neutron star TOV mass, allowing constraints on the characteristic mass to be directly mapped to {upper limits on} the TOV mass. This establishes a novel, independent method for {constraining} the neutron star TOV mass and their equation of state using gravitational-wave and gamma-ray burst observations.} 

\end{abstract}

 \section{Introduction}
The connection between short-duration gamma-ray bursts (sGRBs), kilonova, and binary neutron star (BNS) mergers has been strongly established after the joint detection of the gravitational-wave (GW) event GW170817, the sGRB GRB 170817A, and kilonova AT2017gfo~\citep{LIGOScientific:2017zic,LIGOScientific:2017ync,2017Sci...358.1556C}. By contrast, long-duration gamma-ray bursts (lGRBs) were thought to be solely associated with the collapse of massive stars. However, recent discoveries of lGRB, GRB 211211A~\citep{2022Natur.612..223R,2022Natur.612..228T,2022Natur.612..232Y,2022ApJ...933L..22Z} and GRB 230307A~\citep{2024Natur.626..737L,2025NSRev..12..401S,2024Natur.626..742Y}, associated with kilonova-like emissions, suggested their potential connection with BNS or neutron star--black hole (NSBH) mergers. If BNSs or NSBHs produce both sGRBs and kilonova-associated lGRBs, it raises the questions of what determines the GRB duration and what is the nature of their central engines~\citep{2025ApJ...984...77G,2023ApJ...958L..33G,2025arXiv250510165K,2025PhRvD.111b3010R,2025ApJ...979..190R}. 

BNS and NSBH mergers could produce black holes (BHs) or neutron stars (NSs, in the case of BNSs) as the central engines that power GRBs (e.g.,\citet{1993Natur.361..236M,1999ApJ...527L..39J,2005A&A...436..273A,2011ApJ...732L...6R,2012MNRAS.419.1537B}). The outcomes of the mergers, such as the lifetime of the remnant NSs and the properties of the disk formed around the systems, depend on the mass, spin, spin-orbit alignment, and nuclear-matter properties of the stars. For example, heavier binary systems tend to collapse into BHs on a shorter timescale. 
Recent comparisons with numerical simulations indicated that the duration of the GRBs is likely highly correlated with the mass of the binaries. Heavier binary systems form BHs shortly after mergers and power lGRBs, while lighter systems maintain longer-lived NSs and power sGRBs~\citep{2023ApJ...958L..33G,2025ApJ...984...77G}.

\citet{2023ApJ...958L..33G} found that the duration of BH-powered jets is governed by the timescale over which the BH disk becomes magnetically arrested \citep[MAD;][]{2003PASJ...55L..69N,2011MNRAS.418L..79T}. Numerical simulations show that BH disks with mass $ M \gtrsim 0.01-0.1\,M_\odot $ consistently produce lGRBs with typical GRB luminosities \citep{2023ApJ...954L..21G,2023ApJ...958L..33G}. To investigate whether less massive BH disks produce either fainter lGRBs or standard sGRBs, \citet{2025arXiv250113154I} conducted a parameter study of BH disks and the magnetic flux on the BH. They found that, regardless of disk mass, the jets are likely to be of long duration, but their luminosity scales with the BH disk mass. This implies that lighter disks produce similarly long-duration GRBs, but with luminosities too low to match those of typical GRBs. Consequently, NSs are naturally favored as the central engines of standard sGRBs. This conclusion is further supported by the observed correlation between lGRBs and red kilonovae, and sGRBs with bluer kilonovae \citep{2025ApJ...984...77G}, the latter being indicative of a longer-lived remnant.

Therefore, light BNS mergers that form long-lived NS remnants (but not stable NSs, see Sections~\ref{sec:method} and~\ref{sec:model} below) can generate sGRBs. In contrast, BHs with sufficiently massive accretion disks produce the observed lGRB population. Such massive BH disks arise either from binary mergers with specific mass ratios, e.g., BNS with mass ratio $ q \gtrsim 1.2 $ or NSBH with $ q \lesssim 3 $ \citep[e.g.,][]{2006PhRvD..73f4027S,2012PhRvD..86l4007F,2019ARNPS..69...41S}, or from short-lived NS remnants that enrich their surrounding accretion disks via neutrino-driven ejecta before collapsing into a BH \citep[e.g.,][]{2018ApJ...869..130R}. Both scenarios are associated with heavier mergers than sGRB progenitors.

{
The lifetime of NS merger remnants is strongly correlated with both the mass of the system and the equation of state (EoS) of NSs. For example, if the total mass of the binary exceeds a critical threshold, the remnant collapses to a BH immediately after the merger. This threshold is known as the \textit{prompt-collapse mass}. Numerical studies~\citep{2005PhRvD..71h4021S,2011PhRvD..83l4008H,2013PhRvL.111m1101B} have demonstrated that the prompt-collapse mass is proportional to the maximum mass of cold, non-rotating NSs, also known as the Tolman–Oppenheimer–Volkoff (TOV) mass. The NS TOV mass is a fundamental property that is determined by the NS EoS. Determining the TOV mass will place constraints on the EoS. Extensive theoretical work has been devoted to exploring the TOV and prompt-collapse masses~\citep{1996ApJ...470L..61K,2012ARNPS..62..485L,2013PhRvL.111m1101B,2017PhRvD..96l3012S,2021ApJ...922L..19T,2022PhRvD.106d4026K,2022MNRAS.513.3646P,2022PhRvL.129c2701P,2022PhRvD.105j3022K,2022ApJ...939...51M,2024arXiv240211013E,2024MNRAS.527.8043C,2024PhRvD.109l3011S,2024LRR....27....3K}, while observationally constraining them remains challenging~\citep{2015ApJ...808..186L,2017ApJ...850L..19M,2018ApJ...852L..25R,2018ApJ...858...74M,2018MNRAS.478.1377A,2020PhRvD.102f4063C,2020PhRvD.101f3029S,2024PhRvD.109d3052F,2024arXiv240711153C,2024PhRvD.109d4071D,2025ApJ...987...56L,2025PhRvX..15b1014K}. {Previous studies of sGRB have explored the possibility of constraining the NS EoS using the quasiperiodic oscillation frequency~\citep{2025ApJ...983...88G,2025ApJ...980..220H} and the duration of the X-ray afterglow plateau~\citep{2025ApJ...993...68M} observed with sGRBs.}

In this paper, we investigate additional mass scales that characterize the lifetime of NS merger remnants (see Section~\ref{sec:method} for detailed descriptions). Our goal is to examine the relationship between these characteristic masses and the TOV mass by connecting GW to GRB observations. Establishing such a relation would allow observations that {measure} the characteristic masses to be used to {constrain} the TOV mass, as we demonstrate at the end of this paper.
} 

LIGO-Virgo-KAGRA (LVK) have been observing BNS and NSBH mergers in gravitational waves and place constraints on the merger rates for binaries with different component masses~\citep{KAGRA:2021duu,LIGOScientific:2024elc}. On the other hand, estimates of the rates of sGRB and kilonova-associated lGRB are available from observations~\citep{2012MNRAS.425.2668C,2018Galax...6..130M,2023ApJ...959...13R,2024ApJ...976L..10K,2024arXiv241117244H,2022Natur.612..228T,2024Natur.626..737L}. If the division between sGRBs and lGRBs is related to the mass of the binary, comparison between the GW-observed merger rates at different masses and the GRB rates will reveal these characteristic division masses and potentially placing constraints on the NS EoS that determine the outcome of the mergers. 

Previous work assumed the NS mass distribution follows the Galactic NS observations and identified a transition mass for the durations of the BNS remnants~\citep{2024arXiv241207846P}. In this paper, instead of the Galactic NS mass distribution, which may be different from the cosmic NS mass distribution in mergers~\citep{2021ApJ...921L..25L}, we directly use the LVK observed population~\citep{KAGRA:2021duu}. In addition, with the LVK observations, we are able to include NSBHs as the potential progenitors of GRBs by considering the NS and BH population distribution altogether~\citep{LIGOScientific:2024elc}. We develop a Bayesian framework to investigate the correlations between different {characteristic masses} and properly consider the observational and theoretical uncertainties. 

\section{Method}\label{sec:method}
There are several characteristic total masses ($\mtot=m_1+m_2$) of the binary system that are relevant to the lifetime of the merger remnants. These characteristic masses depend on the nuclear matter properties of the NSs, and they can be correlated with each other. Therefore, we start with one of the fundamental masses, the {NS TOV mass} ($M_{\rm TOV}$), and parameterize other characteristic masses relative to it as follows:
\begin{itemize}
    \item $M_{\rm VL}\equiv k_1 \mtov$, \textit{very long-lived/long-lived remnant}: $M_{\rm VL}$ is the maximum mass of a uniformly rotating NS. If $\mtov < \mtot \leq M_{\rm VL}$, the merger forms a very long-lived NS that lasts for $ t \gg 1\,{\rm s} $ before collapse. Such very long-lived NSs would inject an enormous amount of energy, which is not observed, suggesting that such remnants may not exist in nature \citep[see e.g.,][]{2021ApJ...920..109B}. Notably, this timescale also corresponds to the typical upper limit of sGRB durations.

    \item $M_{\rm LS}\equiv k_2 \mtov$, \textit{long-lived/short-lived remnant}: If $M_{\rm VL}< \mtot \leq M_{\rm LS}$, the system collapses on the viscous timescale that lasts for hundreds of ms. This timescale is governed by angular momentum transport processes \citep{2013PhRvD..88d4026H,2014ApJ...790...19K} or neutrino \citep{2011PhRvL.107e1102S}, and is characteristic of sGRBs.

    \item $M_{\rm PC}\equiv k_3 M_{\rm TOV}$, \textit{short-lived remnant/prompt collapse}:  If $M_{\rm LS}< \mtot \leq M_{\rm PC}$, the system collapses on the GW timescale of $ \sim 10\,{\rm ms} $, too brief for a sGRB, yet long enough to enrich the accretion disk with sufficient mass for the newborn BH to produce a standard lGRB. If $\mtot>M_{\rm PC}$, the system collapses dynamically into a BH within several ms after the merger, in which case the disk mass is determined by the binary mass ratio.
\end{itemize}
It follows from the above definitions that $k_3>k_2>k_1>1$. We then infer $(k_1,k_2,k_3,M_{\rm TOV})$ by comparing GW and GRB observations {as described below.}

{We use the LVK's GWTC-4 catalog~\citep{LIGOScientific:2025slb} for the local ($z\lesssim 0.2$) merger rate as a function of binary masses. Specifically, we adopt the \texttt{FullPop-4.0} population model from~\citet{LIGOScientific:2025pvj}; other population model choices in that study predict consistent merger rates. We randomly draw 1000 rate posterior samples provided by~\citet{LIGOScientific:2025slb,LIGOScientific:2025pvj} data release. }Each sample gives a local merger rate estimate for mergers with component mass $(m_1,m_2)$. 

On the other hand, the observations of sGRBs and kilonova-associated lGRBs have placed constraints on their rates~\citep{2012MNRAS.425.2668C,2018Galax...6..130M,2023ApJ...959...13R,2024ApJ...976L..10K,2024arXiv241117244H,2022Natur.612..228T,2024Natur.626..737L}
The intrinsic rate of sGRBs, $R_{\rm sGRB}$, lies between $O(10)$ to 2000 Gpc$^{-3}$yr$^{-1}$. Given the uncertainty in kilonova-associated lGRB rates{, $\rlgrb$,} and the wide range of values reported in the literature, the rate ratio between the two types of GRBs, defined as $\eta \equiv R_{\rm lGRB}/R_{\rm sGRB}$, spans a broad range from 0.5\% to 100\%.

In order to compare the GW merger rates to GRB rates, we follow the definitions of characteristic total masses above and consider mergers with $M_{\rm VL}\leq \mtot< M_{\rm LS}$ as sGRBs. If $M_{\rm LS} \leq \mtot< M_{\rm PC}$, we consider them as lGRBs. {Mergers with $\mtot\geq M_{\rm PC}$ that undergo prompt collapse to a BH leave substantial mass outside the BH innermost stable circular orbit, forming an accretion disk capable of powering a lGRB, provided their mass ratios ($q=m_1/m_2$) fall within $q_{\rm low}\leq q \leq q_{\rm high}$, where $q_{\rm low}=1.2$ and $q_{\rm high}=3$. In prompt collapse BNS mergers with $q < q_{\rm low} $, the near-symmetric masses suppress tidal torques and tails, so most material has insufficient specific angular momentum and plunges into the BH on a dynamical time, precluding disk formation. In NSBH mergers with $ q > q_{\rm high} $, the NS is swallowed before tidal disruption outside the last stable orbit, likewise preventing a disk. Mergers with other masses are considered to power either no GRB or low-luminosity GRBs, which are not considered in our comparison. \textit{Although we start with these values for our fiducial model, we relax them and discuss the implications later} (see Section~\ref{sec:model} and Table~\ref{tab:ktable}).} Following the above criteria, we add up the GW merger rates in the corresponding mass ranges to obtain the rates for merger-induced sGRBs, $\rgwsgrb$, and merger-induced lGRBs, $\rgwlgrb$.

With the GRB and GW observational data, $\cal{D}_{\rm GRB}$ and $\cal{D}_{\rm GW}$, the set of {merger characteristic} parameters $\vec\Theta\equiv(k_1,k_2,k_3,M_{\rm TOV})$ can be inferred following the Bayes' theorem:  

\begin{equation}\label{eq:bayes}
    P(\vec\Theta|\mathcal{D}_{\rm GRB},\mathcal{D}_{\rm GW})=\frac{P(\vec\Theta) \times P(\mathcal{D}_{\rm GRB},\mathcal{D}_{\rm GW}|\vec\Theta)}{P(\mathcal{D}_{\rm GRB},\mathcal{D}_{\rm GW})}.
\end{equation}
For the prior $P(\vec\Theta)$, we begin with a uniform prior on $k_1$ and $k_2$ and require $k_3>k_2>k_1>1$. $k_3$ specifies the prompt-collapse mass for NSs, which has been widely explored in numerical simulations~\citep{2013PhRvL.111m1101B,2017ApJ...850L..34B,2021ApJ...922L..19T,2022PhRvD.105j3022K,2022PhRvL.129c2701P}. We adopt a semi-analytical fitting formula and their coefficients from Equation 1b of ~\citet{2022PhRvD.105j3022K}, $M_{\rm PC}(R_{\rm TOV},\mtov)$, where $R_{\rm TOV}$ represents the radius of a NS at its $\mtov$ mass. This formula connects the NS prompt-collapse mass $\mpc$ to its TOV radius and mass. We use the public NS radius and mass posterior samples of \citet{2021PhRvD.104f3003L} as the prior for $R_{\rm TOV}$ and $\mtov$. {Specifically, we use the \texttt{LCEHL\_EOS\_posterior\_samples\_PSR+GW+NICER.h5} file that conditioned on the GW observations~\citep{LIGOScientific:2018hze,LIGOScientific:2020aai}, PSR J0348+0432 mass measurement~\citep{2013Sci...340..448A},  mass and radius measurements for PSR J0030+0451~\citep{2019ApJ...887L..24M}, and mass and radius measurements for PSR J0740+6620~\citep{2021ApJ...918L..28M}.} The {NS radius and mass} constraints encoded in this posterior distribution are broadly consistent with those of e.g.~\citet{2022Natur.606..276H,2020NatAs...4..625C}, which include additional astrophysical and nuclear-physics constraints.

The likelihood function {in Eq.~\ref{eq:bayes}}, $P(\mathcal{D}_{\rm GRB},\mathcal{D}_{\rm GW}|\vec{\Theta})$, represents the probability of the GRB and GW observations given the {characteristic} parameters $\vec{\Theta}$. In our inference, this is the probability of having \textit{observed} sGRB rate $\rsgrb$, and rate ratio $\eta$, assuming the GRBs were produced from the GW observed merger population following the mass criteria described above, or, 
\begin{align}\label{eq:grblikelihood}
    & P(\rsgrb,\eta|\vec\Theta) \nonumber \\
    & =\int f(\rsgrb,\eta|\rgwsgrb,\eta_{\rm GW},\vec{\Theta})\times \nonumber \\ & f(\rgwsgrb,\eta_{\rm GW}|\vec{\Theta})\,d\rgwsgrb \,d\eta_{\rm GW},
\end{align}
where $\eta_{GW}\equiv {\rgwlgrb}/{\rgwsgrb}$. $f(a|b)$ represents the conditional probability density function of $a$ given $b$. The distribution of $f(\rgwsgrb,\eta_{\rm GW}|\vec{\Theta})$ is given by the observed GW merger rate for any given $\vec{\Theta}$ as described above. Given the large uncertainties of GRB observed rates, we use two independent normal distributions to model
\begin{align}\label{eq:grbgwlikelihood}
&f(\rsgrb,\eta|\rgwsgrb,\eta_{\rm GW},\vec{\Theta}) \nonumber \\
& =\frac{1}{\sqrt{2\pi}\sigma_{\rsgrb}}{\rm exp}\{-\frac{[\rsgrb-\rgwsgrb(\vec{\Theta})]^2}{2\sigma^2_{\rsgrb}}\} \times \nonumber \\
& \frac{1}{\sqrt{2\pi}\sigma_{\eta}}{\rm exp}\{-\frac{[\eta-\eta_{\rm GW}(\vec{\Theta})]^2}{2\sigma^2_{\eta}}\}. 
\end{align}
$\rgwsgrb(\vec{\Theta})$ and $\eta_{\rm GW}(\vec{\Theta})$ are the values of $\rgwsgrb$ and $\eta_{\rm GW}$ with given $\vec{\Theta}$. We start with $(\rsgrb,\sigma_{\rsgrb})=(100,100)$ Gpc$^{-3}$yr$^{-1}$ and $(\eta,\sigma_{\eta})=(1,1)$, and explore other possible values below.

\section{Result}\label{sec:result}

\begin{figure}
    \centering
    \includegraphics[width=1.0\linewidth]{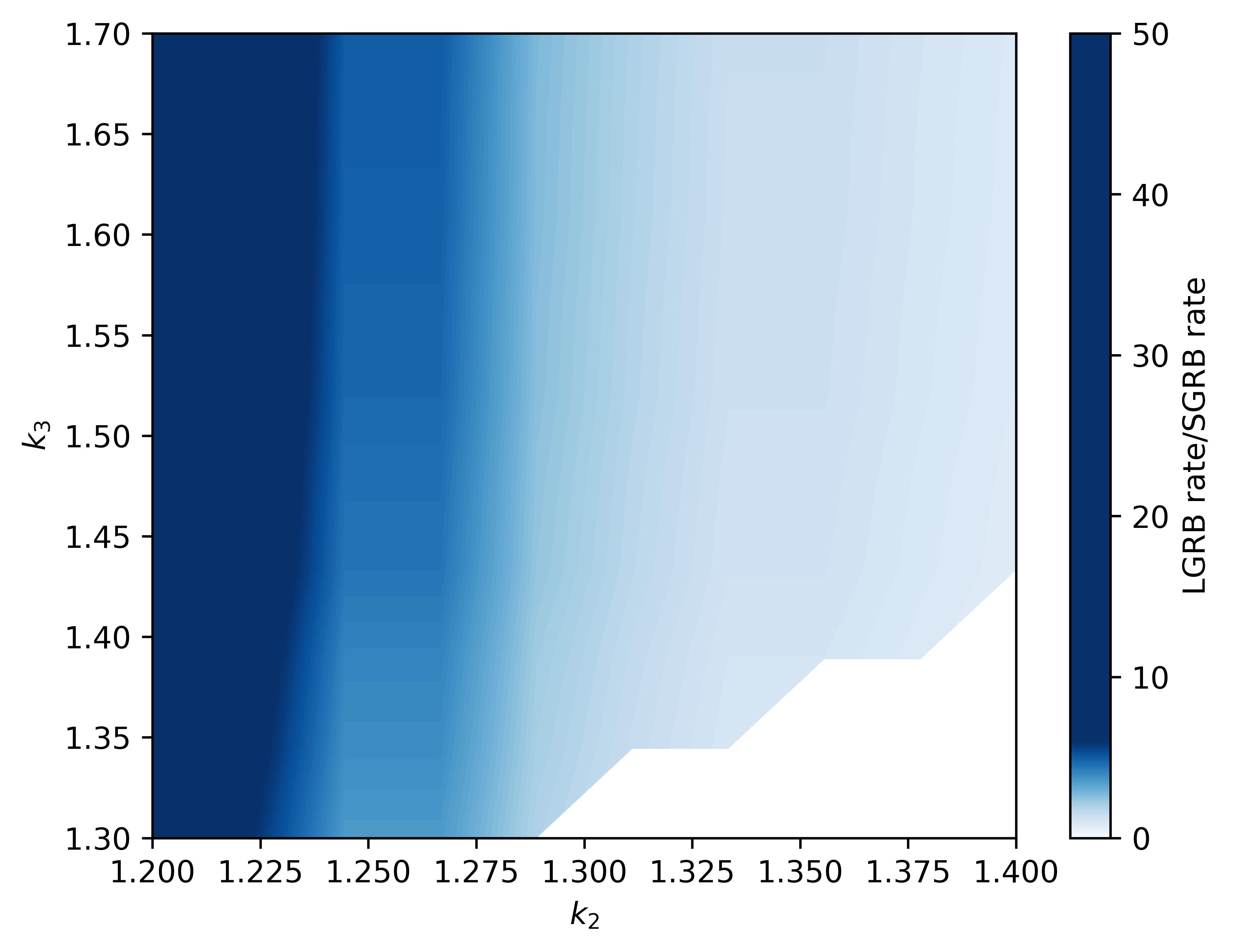}
    \caption{
    {GRB rate ratio shown in the $(k_2,k_3)$ plane for fixed $k_1=1.15$ and $M_{\rm TOV}=2.2\,M_{\odot}$. The rate ratio is highly sensitive to $k_2$, varying by more than the current observational uncertainty over much of the parameter space. This suggests that GRB rate measurements can place informative constraints on $k_2$.}
    } 
    \label{fig:rateratio}
\end{figure}

{In Figure~\ref{fig:rateratio}, we illustrate how $(k_2,k_3)$ affect the predicted rates. Fixing $k_1=1.15$ and $M_{\rm TOV}=2.2\,M_{\odot}$, we show the merger-induced GRB rate ratio, $\eta_{\rm GW}(\vec{\Theta})$, across the two-dimensional $(k_2,k_3)$ parameter space. We find that $\eta_{\rm GW}$ is highly sensitive to $k_2$, with substantial variations arising from relatively small changes in this parameter. The resulting spread in the predicted rate ratio is considerably larger than the current observational uncertainty. This suggests that GRB rate measurements can provide meaningful constraints on $k_2$ despite their present uncertainties.}

\begin{figure*}
    \centering
    \includegraphics[width=0.8\linewidth]{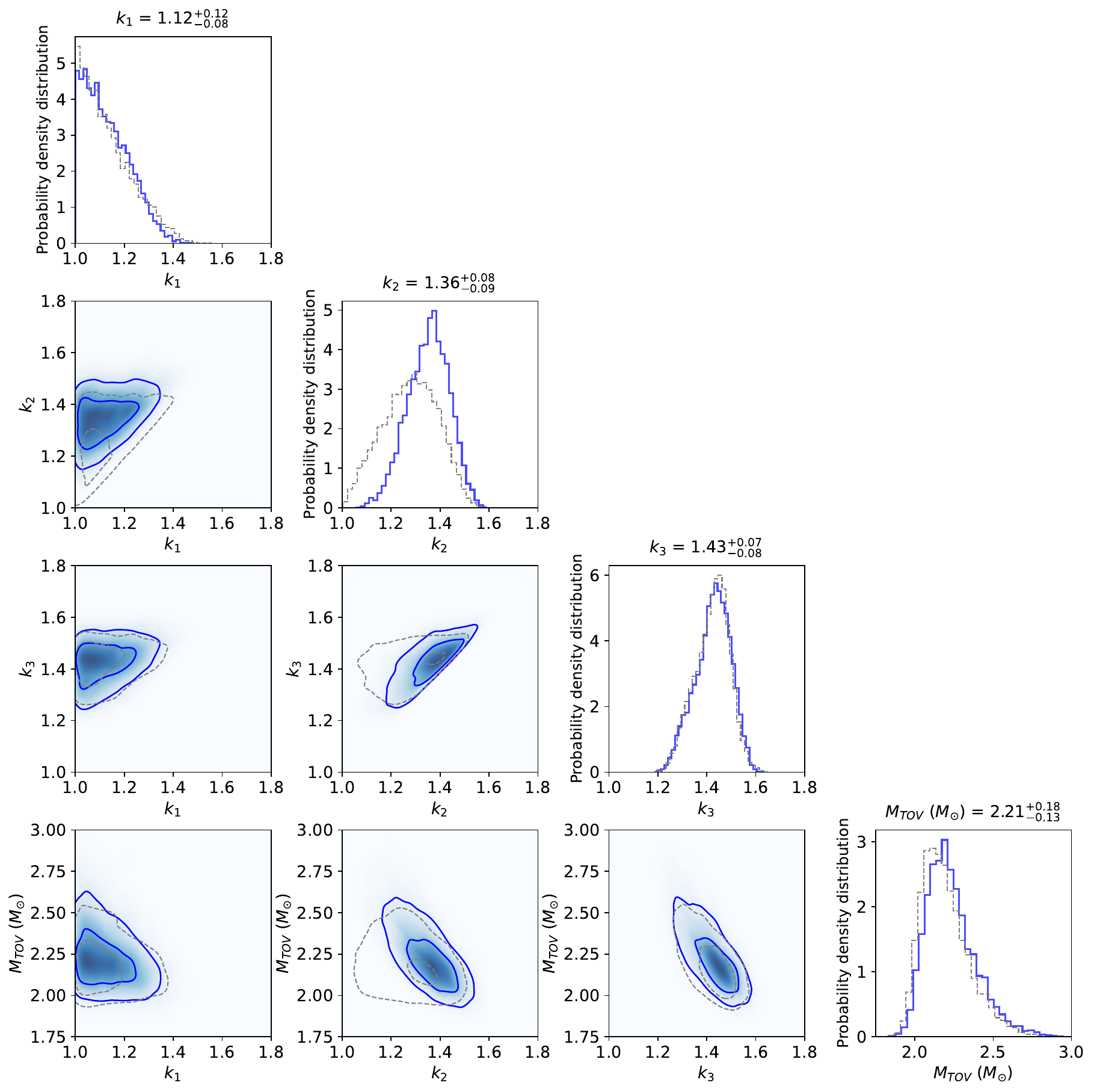}
    \caption{Marginalized 1- and 2-dimensional posteriors of the $(k_1,k_2,k_3,\mtov)$ inference with the fiducial parameters (see Table~\ref{tab:ktable}). The 50\% and 90\% confidence intervals are denoted as contours. The priors are shown in gray dashed lines (we note that although the priors on $k_1$ and $k_2$ are uniform, the  $k_3>k_2>k_1>1$ requirement makes them no longer flat.)}
    \label{fig:corner_inf}
\end{figure*}
In Figure~\ref{fig:corner_inf}, we show the marginalized 1- and 2-dimensional posteriors of the $(k_1,k_2,k_3,\mtov)$ inference, and we summarize the inferred values in Table~\ref{tab:ktable} (``Fiducial parameters''). Compared to the prior (grey dashed lines), the parameter that determines the division between long-lived and short-lived remnant, $k_2$, is the most informative. In particular, we find a relatively high $k_2=1.36^{+0.08}_{-0.09}$ (median and 68\% confidence interval), or $k_2>1.24$ with 90\% confidence. Large $k_2$ indicates that heavy NSs could survive for a relatively long timescale after merger. 

 We also find strong correlation between $(k_2, k_3)$ and anti-correlation between $(k_2,\mtov)$. Both $k_2$ and $k_3$ influence the lGRB population. As $k_2$ increases, $k_3$ must also increase to maintain a consistent event rate. The anti-correlation between $k_2$ and $\mtov$ arises from a prior-imposed anti-correlation between $k_3$ and $\mtov$ (see {the prior contour of $k_3$ and $\mtov$ in} Fig.~\ref{fig:corner_inf}). {Through this anti-correlation, a determination of $k_2$ can be converted into constraints on  $\mtov$.}

As an exploratory test, we reduce the uncertainty in the GRB rate and the GW rate parameters by an order of magnitude, and find that the correlation between $(k_2, k_3)$ and anti-correlation between $(k_2,\mtov)$ are tightened. {Therefore, future improvements in GRB and GW merger rate measurements are expected to provide a more precise mapping between $k_2$, $k_3$, and $\mtov$.}

\begin{table*}[]
    \centering
    \begin{tabular}{c|c|c|c|c|c}
        Variations & $k_1$ & $k_2$ & $k_3$ & $M_{\rm TOV}$ \\ \hline
        Fiducial parameters& $1.12^{+0.12}_{-0.08}$ & $1.36^{+0.08}_{-0.09}$ & $1.43^{+0.07}_{-0.08}$ & $2.21^{+0.18}_{-0.13}$\\ 
        $R_{\rm sGRB}=30^{+30}_{-30}$ Gpc$^{-1}$yr$^{-1}$ & $1.14^{+0.11}_{-0.1}$ & $1.34^{+0.08}_{-0.09}$ & $1.42^{+0.06}_{-0.09}$ & $2.22^{+0.2}_{-0.14}$\\
        $R_{\rm sGRB}=1000^{+500}_{-500}$ Gpc$^{-1}$yr$^{-1}$ & $1.11^{+0.11}_{-0.07}$ & $1.35^{+0.08}_{-0.09}$ & $1.43^{+0.06}_{-0.09}$ & $2.21^{+0.2}_{-0.13}$\\
        $R_{\rm lGRB}/R_{\rm sGRB}=0.01^{+0.01}_{-0.01}$ & $1.07^{+0.08}_{-0.05}$ & $1.33^{+0.09}_{-0.09}$ & $1.44^{+0.06}_{-0.08}$ & $2.15^{+0.13}_{-0.11}$ \\
        $R_{\rm lGRB}/R_{\rm sGRB}=3^{+1}_{-1}$ & $1.15^{+0.11}_{-0.1}$ & $1.31^{+0.09}_{-0.09}$ & $1.41^{+0.07}_{-0.09}$ & $2.21^{+0.21}_{-0.14}$\\
        $q_{\rm low}=1.15$ & $1.11^{+0.11}_{-0.08}$ & $1.35^{+0.08}_{-0.1}$ & $1.43^{+0.06}_{-0.09}$ & $2.21^{+0.2}_{-0.14}$  \\
        $q_{\rm low}=1.35$ & $1.12^{+0.12}_{-0.09}$ & $1.35^{+0.08}_{-0.09}$ & $1.43^{+0.06}_{-0.08}$ & $2.2^{+0.2}_{-0.13}$  \\ 
        $q_{\rm high}=2$ & $1.12^{+0.11}_{-0.09}$ & $1.35^{+0.08}_{-0.1}$ & $1.43^{+0.06}_{-0.09}$ & $2.2^{+0.21}_{-0.13}$\\        
        $q_{\rm high}=10$ & $1.11^{+0.11}_{-0.07}$ & $1.35^{+0.08}_{-0.09}$ & $1.43^{+0.06}_{-0.08}$ & $2.2^{+0.2}_{-0.12}$  \\
        Without $\mvl$ & N/A & $1.3^{+0.12}_{-0.12}$ & $1.39^{+0.09}_{-0.11}$ & $2.29^{+0.28}_{-0.18}$    
    \end{tabular}
    \caption{Median and 68\% confidence interval of the inferred $(k_1,k_2,k_3,\mtov)$ values, shown for various observed GRB rates and choices of mass criteria. The fiducial parameters are $R_{\rm sGRB}=100^{100}_{-100}$ Gpc$^{-1}$yr$^{-1}$, $R_{\rm lGRB}/R_{\rm sGRB}=1^{+1}_{-1}$, $q_{\rm low}=1.2 $, and $q_{\rm high}=3$.}
    \label{tab:ktable}
\end{table*}

\subsection{Robust against observational uncertainties}\label{sec:observation}
In Table~\ref{tab:ktable}, we explore how observational uncertainties affect our results. Although the GRB observed rate remains uncertain and modeled using normal distributions to capture the uncertainties (Eq.~\ref{eq:grbgwlikelihood}), the inference of $k_2$ remains largely robust across a broad range of assumptions, from $\rsgrb = 30$ to $2000$ Gpc$^{-3}$yr$^{-1}$ and $\eta$ values spanning 0.01 to 3 (see Table~\ref{tab:ktable}). These variations mainly affect $k_1$ and $\mtov${, instead of $k_2$,} due to the steeper slope of the total mass distribution near and below $\mvl$ in the GW observations. Modifications to $k_1$ and $\mtov$ have a stronger impact on the sGRB rate and the rate ratio.

{On the other hand, }the observed GW merger rate uncertainty is already incorporated in our results (Eq.~\ref{eq:grblikelihood}).

\subsection{Agnostic to model}\label{sec:model}
We explore how the choice of {mass} criteria affects the results by varying the parameters that were chosen, i.e., parameters other than $(k_1,k_2,k_3,\mtov)$. 

Considering uncertainties in the mass ratios needed to induce significant torque on a deformed NS in BNS mergers, we allow the lower mass ratio threshold, $q_{\rm low}$ to range between 1.15 to 1.35 \citep{2019ARNPS..69...41S}. Similarly, we permit variation in the mass ratio required for substantial disk formation following NSBH mergers, with the upper threshold, $q_{\rm high}$ ranging from 2 to 10 \citep{2012PhRvD..86l4007F}. When we add up the GW merger rates within the mass range of interest, we find these choices of cut in mass ratio only affect less than 20\% of mergers with $\mtot>\mtov$. As a result, different choices {of mass ratio threshold} have no substantial impact (see Table~\ref{tab:ktable}).

NS remnants with $ M < \mvl $ would form a very long-lived NS. Although recent studies have proposed magnetars as possible progenitors of kilonova-associated lGRBs \citep{2022Natur.612..232Y,2025NSRev..12..401S,2025ApJ...978L..38C}, observational evidence strongly disfavors magnetar-driven GRB scenarios due to the lack of significant rotational energy injection signatures \citep{{2014MNRAS.437.1821M,2016ApJ...819L..22H,2020ApJ...902...82S,2021ApJ...920..109B}}. Therefore, we have assumed a minimum total mass $\mvl$ corresponding to the minimum mass for forming a meta-stable NS that would produce a sGRB.
The fraction of GW mergers with $\mtot<\mvl$ (say, if we assume $\mtov=2.2\msun$ and $k_1=1.15$) is less than 20\%. If we completely remove the restriction for a minimum total mass and assume all mergers with $\mtot<\mls$ are sGRBs, we do not find significant difference (see Table~\ref{tab:ktable}). 

In summary, although the choices of mass cuts at the boundaries (e.g., light systems, heavy systems with low or high mass ratio) have not yet been fully explored theoretically, our results appear agnostic.

 \section{Discussion}
In this paper, we introduce a new approach to {constrain} the NS TOV mass by comparing the GW and GRB observations. Our method provides an independent {upper limit} on the NS TOV mass, complementing existing techniques and potentially offering tighter limits on the NS EoS with future GW and GRB observations. Our findings appear to be insensitive to the large observational uncertainties and agnostic to other mass-related assumptions. {However, we expect both observational uncertainties and model assumptions to become increasingly important as observational constraints improve and the measurements become more precise.}

Interestingly, \citet{2024arXiv241207846P} found the short- to long-lived remnant transition to occur between 1.3 and 1.4 $\mtov$ in numerical relativity simulations. Our analysis, based entirely on GW and GRB observations, arrives at consistent conclusions. These two independent studies both point to a long post-merger lifetime for relatively massive NSs.

In our analysis, we find a strong correlation between $(k_2,k_3)$ and an anti-correlation between $(k_2,\mtov)$. Larger $\mls$ indicates smaller NS TOV mass $\mtov$ and larger prompt collapse mass $\mpc$. 
Currently, the Kullback–Leibler (KL) divergence in our analysis, which evaluates the posterior-to-prior volume ratio, is around 0.5.
Improved rate measurements of GW and GRB events in the future will  help further refine the correlations.
Future joint detections of GRBs with BNS or NSBH mergers in GWs will link the total merger mass to the GRB duration, yielding an estimate of $M_{\rm LS}$ and/or $\mpc$. These estimates can also be used to tighten their correlation with $\mtov$.

If we conservatively assume that the uncertainty on the rate will be reduced by a factor of 3 at the end of the LVK fifth observing run (O5), a single long GRB-GW joint detection could reduce the uncertainty on $M_{\rm TOV}$ to $\sim 0.125\,M_{\odot}$. With two long GRB-GW joint detections, the uncertainty could be further reduced to $\sim 0.1\,M_{\odot}$, surpassing several currently available constraints.

Our Bayesian framework ensures proper consideration of observational and theoretical uncertainties and reveals correlations between different {merger characteristic masses}. Although we focus on the total mass in this paper, our framework could also be extended to include other schemes, such as the dependency on the merger mass ratio.

\section*{Acknowledgments}
The authors would like to thank François Foucart, Rosalba Perna, David Radice, and Aaron Zimmerman for useful discussions.
H.-Y.C. is supported by the National Science Foundation under Grant PHY-2308752 and Department of Energy Grant DE-SC0025296.
O.G. is supported by the Flatiron Research Fellowship. The Flatiron Institute is supported by the Simons Foundation.

\bibliography{ref}
\bibliographystyle{aasjournal}

\end{document}